\documentclass[11pt,twoside]{article}
\usepackage{asp2010}

\resetcounters

\begin{document}
 \title{Dynamics of extended AGB star envelopes}
 \author{Claudia~Dreyer, Michael~Hegmann, and Erwin~Sedlmayr
 \affil{Zentrum f\"ur Astronomie und Astrophysik, Technische Universit\"at Berlin, Berlin, Germany}}

 \begin{abstract}
  The dust formed in extended circumstellar envelopes of long-period variables and Miras has a strong 
  influence on the envelope dynamics. A radiatively driven instability caused by the formation of dust 
  leads to the development of an autonomous dynamics characterised by a set of distinct frequencies. We 
  study the interplay between the envelope's internal dynamics and an external excitation by a pulsating 
  star.
 \end{abstract}

 \section{Introduction}
  Long-period variables (LPVs) and Miras are radially pulsating, highly evolved stars on the asymptotic 
  giant branch (AGB). Their cool, extended atmospheres are excellent sites for the formation of complex 
  molecules and dust particles. The interplay between dust formation and stellar radiation results in 
  circumstellar envelopes (CSEs) generating slow mass loss, which finally enriches the interstellar medium 
  with processed material. The observed light curves exhibit irregularities over intervals of several 
  pulsation cycles, which are at least partially caused by the phenomenon of dust formation.

  To investigate the complex dynamical behaviour of carbon-rich envelopes in detail, CSEs can be considered 
  as nonlinear multi-oscillatory systems, whose eigenfrequencies and normal modes are controlled by the 
  intrinsic timescales of various coupled physical and chemical processes. This is done by applying established 
  methods of non-linear dynamics such as Fourier Analysis and the study of stroboscopic and Poincar\'e maps 
  upon the results of self-consistent model calculations \citep{fleischer_1992}.

 \section{Eigendynamics}
\begin{figure}[!htb]
    \centering
     \includegraphics[angle=0,width=0.43\linewidth]{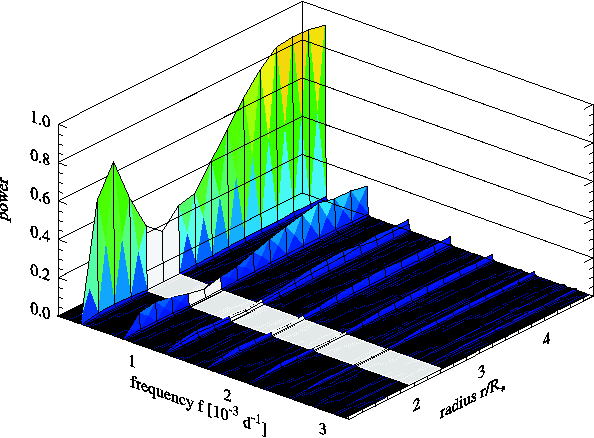}
     \hspace{1cm}
     \includegraphics[angle=0,width=0.43\linewidth]{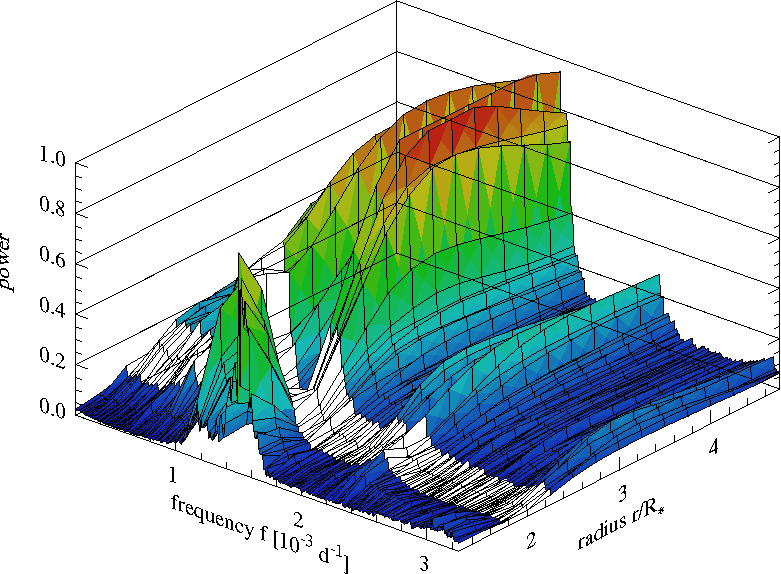}
     \caption{Power spectra of the radial outflow velocity $u$ for a high luminous LPV purely caused by the 
              exterior $\kappa$-mechanism (model A, left) and caused by a stochastic excitation for a standard 
              luminous LPV (model B, right). The position of the dust formation zone in the envelope is 
              highlighted.}
     \label{eigendynamics}
  \end{figure}

  The envelopes around highly luminous AGB stars, e.g.~$L_* = 3\times 10^4\,\mathrm{L_{\odot}}$, $M_* = 1\,
  \mathrm{M_{\odot}}$, $T_* = 2450\,\mathrm{K}$, and $C/O = 1.25$ (henceforth model A) develop a self-maintaining oscillation caused by dust formation (exterior $\kappa$-mechanism) even without the additional 
  input of mechanical momentum by an underlying stellar pulsation \citep[cf.][]{fleischer_1995,hoefner_1995}. 
  In their power spectrum (Fig.~\ref{eigendynamics}, left) one can clearly see the eigenfrequency $f_{\kappa}
  \approx 2450\,\mathrm{d^{-1}}$ and its overtones \citep{dreyer_2009}).

  In the case of lower stellar luminosities, e.g.~$L_* = 9\times10^3\,\mathrm{L_{\odot}}$, $M_* = 1\,
  \mathrm{M_{\odot}}$, $T_* = 2500\,\mathrm{K}$, and $C/O = 1.75$ (henceforth model B), the system generates 
  no self-induced oscillation. 
  An additional input of energy and momentum is needed to determine the CSE eigendynamics. We excite such an 
  envelope by white noise. Thereby we adopt a normal noise distribution with zero mean and intensity $\sigma$ 
  \citep[see][]{dreyer_2010}. The resulting power spectrum (Fig.~\ref{eigendynamics}, right) exhibits different 
  eigenmodes. The position of the minima (marked in grey) coincides with the dust nucleation zone located at 
  $r\approx2.25\,\mathrm{R_*}$. This confirms that the dust is responsible for creating the characteristic 
  frequencies of the envelope.

 \section{Interaction with external excitation}
\begin{figure}[htb]
   \centering
   \includegraphics[width=0.94\textwidth]{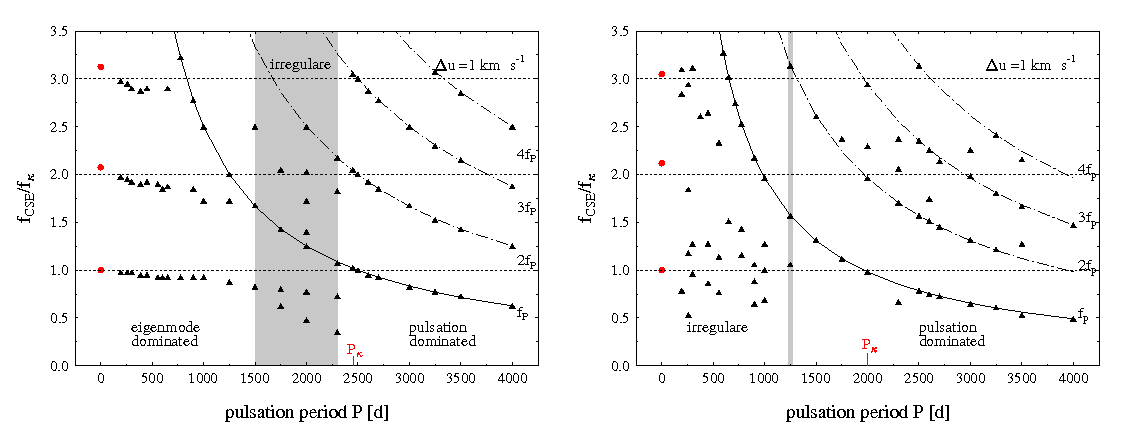}
   \caption{Most dominant frequencies of a CSE $f_{\mathrm{CSE}}$ normalised to its eigenfrequency $f_{\kappa}$  
           (triangles) excited with various pulsation periods $P$ near the dust nucleation zone for model A (left) 
           and model B (right). The set of CSE-eigenmodes is depicted by circles, the eigenperiod is labelled with 
           $P_{\kappa}$. The excitation frequencies (solid line) and harmonics (dash-dotted lines) are also shown.}
   \label{interaction}
  \end{figure}

  Once the eigenfrequencies have been found, the question arises as to how these modes interact with an
  external excitation caused by the stellar pulsation. For this purpose, we studied the response of our 
  reference models for a series of harmonic excitations with different periods P and strength $\Delta u$.
  Figure \ref{interaction} shows the most dominant envelope frequencies over the excitation period at 
  constant strength. For model A (left) we can identify three different dynamic domains, namely eigenmode
  -dominated, irregular and pulsation-dominated domain.
  The same study for model B (right) reveals a similar behaviour. In contrast to model A we have not found 
  an eigenmode dominated domain.

\begin{figure}[!htb]
   \centering
    \includegraphics[width=0.43\textwidth]{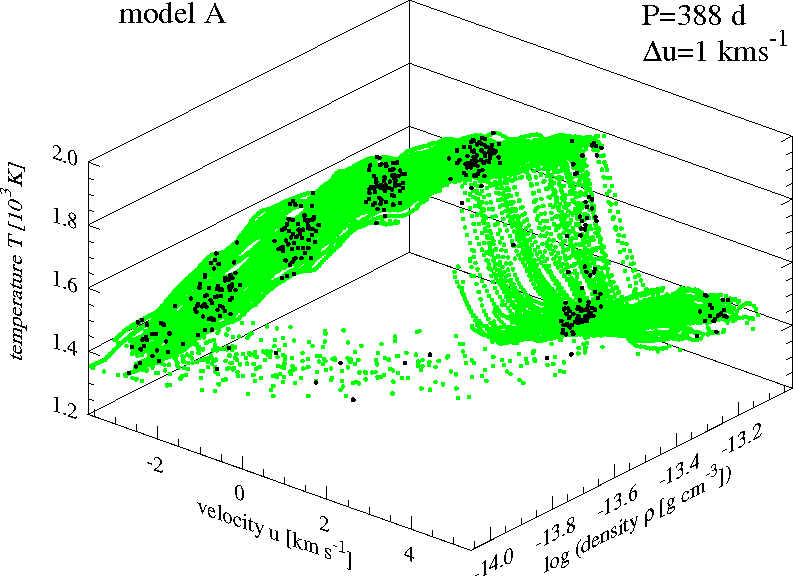}
    \hspace{1cm}
    \includegraphics[width=0.43\textwidth]{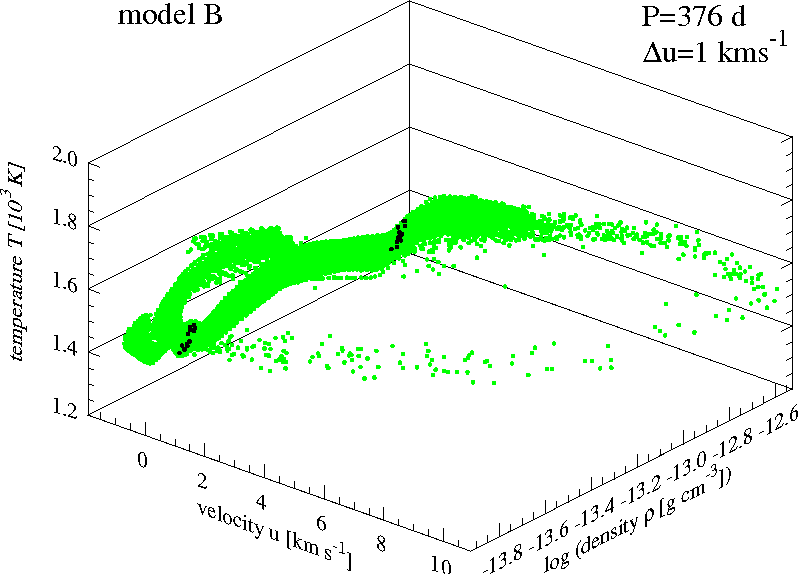}
    \caption{Maps of the ($u$,$\rho$,$T$) phase space in the dust nucleation zone for model A (left) and model B 
            (right). The stroboscopic maps were obtained by sampling $[u(n0.02P), \rho(n0.02P), T(n0.02P)]$ (grey) 
            and the Poincar\'e maps by $[u(nP), \rho(nP), T(nP)]$ (black) for $1\le n \le \lfloor t_{max}/P\rfloor, 
            n\in\mathbb{N}$.}
    \label{resonance}
  \end{figure}

  Figure \ref{resonance} shows the maps of our reference models for a typical oscillation period of LPVs. In the 
  maps it can be seen that for a constant phase angle (black) the system tends to stay inside a finite number of 
  stripes. Similar to the orbital resonance in celestial mechanics, we can find integers $i$ and $j$ which fulfil 
  the resonance condition $iP=jP_{\mathrm{CSE}}$. This suggest that the systems returns to the same physical state 
  after $i$ excitation periods $P$ or $j$ resonance periods $P_{\mathrm{CSE}}$.
  For model A, we find that $i=27$ and $j=4$. For model B the CSE response is double periodic, which means that 
  the envelope period equals two times the excitation period \citep{dreyer_2009,dreyer_2010}.

 \section{Synthetic light curves}
  \begin{figure}[!htb]
   \centering
    \includegraphics[angle=0,width=0.49\textwidth]{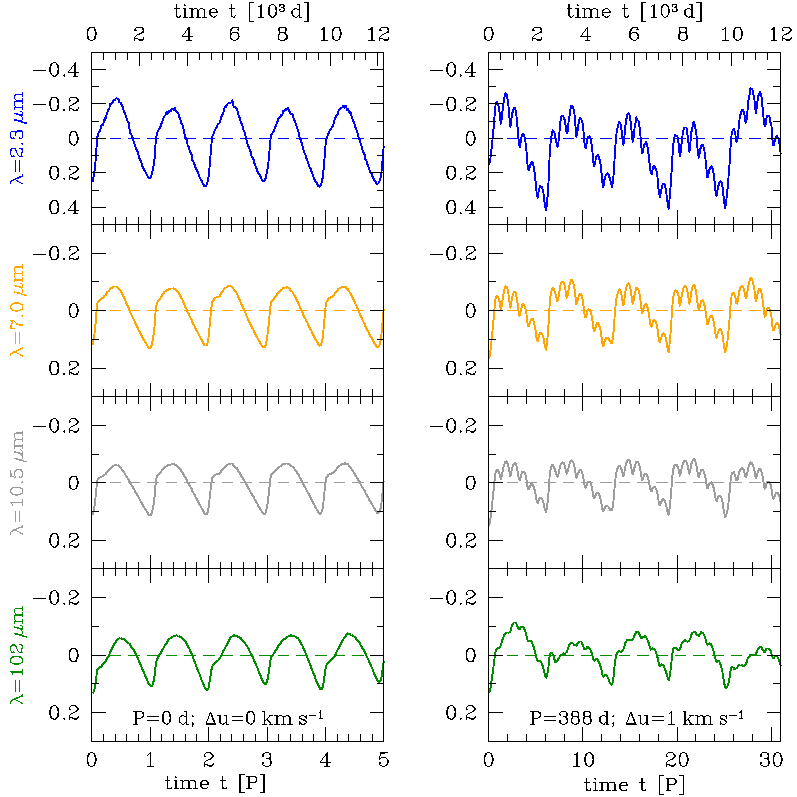} 
    \hfill
    \includegraphics[angle=0,width=0.49\textwidth]{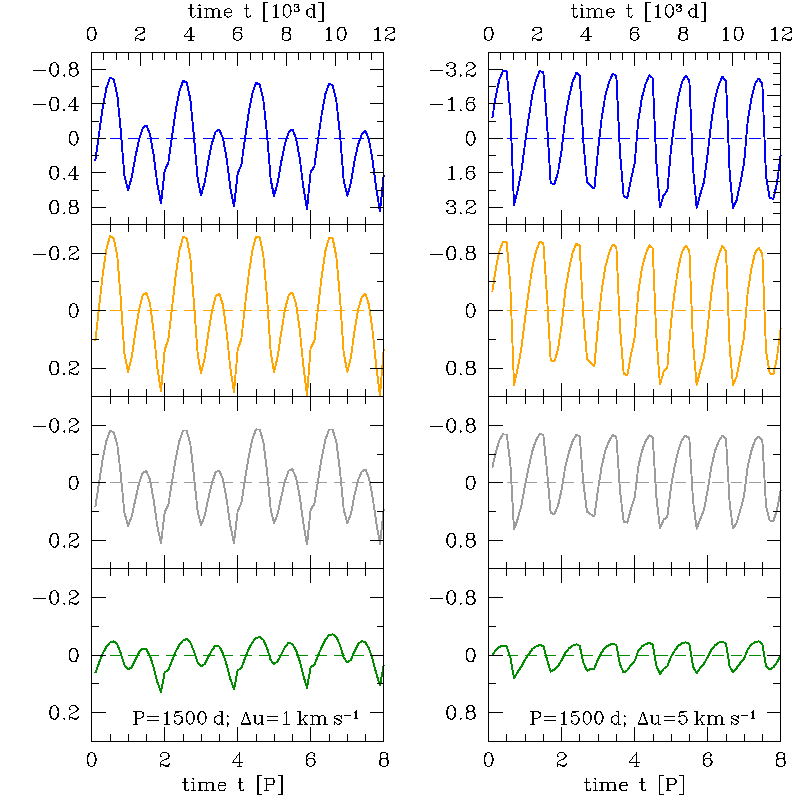} 
    \caption{Synthetic light curves of model A for various excitations at different wavelength. 
             The magnitudes are related to the mean magnitude.}
    \label{lightcurve_a10i}
  \end{figure}
  Figure \ref{lightcurve_a10i} presents synthetic light curves for model A obtained for various excitations 
  at several wavelengths. The light curves in the first and the last panels show monoperiodic behaviour but 
  with differing periods. For the unexcited CDS (left) the dominant timescale is the dust formation scale 
  $P_{\kappa} = 2450\,\mathrm{d}$, whereas in the case of the highly-excited model (right), this timescale 
  is determined by the stellar pulsation $P=1500\,\mathrm{d}$. 
  In the curves in the second panels the eigenperiod of the CSE $P_{\kappa} = 2450\,\mathrm{d}$ and the 
  excitation period $P = 388\,\mathrm{d}$ interfere with each other. Nevertheless, the dominant timescale 
  is the slightly shifted dust formation timescale $P_{\mathrm{CSE}}\approx 2700\,\mathrm{d}$.
  The light curves in the third panels show double-periodic behaviour. The eigenperiod is detuned to the 
  first harmonic of the excitation period $P_{\mathrm{CDS}} = 2P = 3000\,\mathrm{d}$.

 \section{Special application: IRC +10216}
\begin{figure}[htb]
   \centering
    \includegraphics[width=0.43\linewidth]{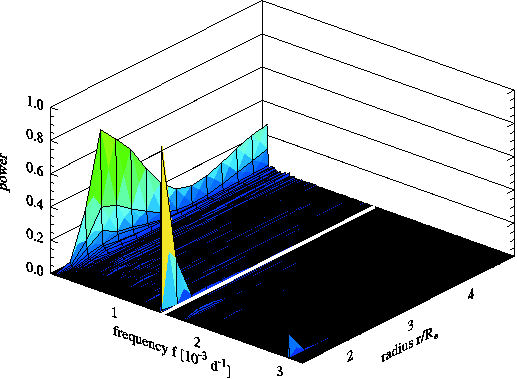}
    \hspace{1cm}
    \includegraphics[width=0.43\linewidth]{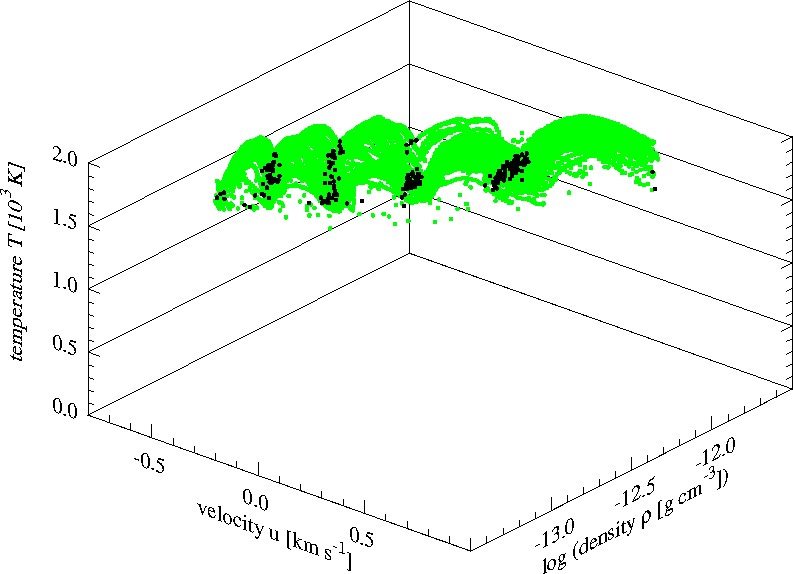}
    \caption{Power spectrum of velocity $u$ (left) and maps of the $(u,\rho,T)$-phase space (right) of the 
             periodically excited IRC +10216 model. The excitation frequency is highlighted in the spectrum.}
    \label{irc}
  \end{figure}

  \citet{winters_1996} have found that the fundamental parameters: $L_* = 24\times10^3\,\mathrm{L_{\odot}}$, 
  $M_* = 1\,\mathrm{M_{\odot}}$, $T_* = 2500\,\mathrm{K}$, $C/O = 1.2$ as well as $P = 650\,\mathrm{d}$, and 
  $\Delta u =6\,\mathrm{km\,s^{-1}}$ well reproduce observations of IRC +10216.
  So, we use this set of parameter to model and investigate the dynamical behaviour of the envelope. To 
  determine the envelope eigendynamics the investigation follows the approach used for CSEs around standard 
  and low luminous AGB stars.
  Since the central star pulsates with a well defined period of $P = 650\,\mathrm{d}$ \citep[e.g.][]{menten_2006}, 
  we excite the model exactly with this frequency. As can be seen in the power spectrum (Fig.~\ref{irc}, left) 
  the CSE is dominated by the pulsation frequency and its first harmonic close to the star.
  With onset of dust formation at $r\approx 2.5\,\mathrm{R_*}$ these frequencies disappear and the envelope 
  frequency $f_{\mathrm{CDS}} = 0.26\times10^{-3} \mathrm{d^{-1}}\approx(3900\,\mathrm{d)^{-1}}$ controlled 
  by the timescale of the dust formation process gains strength. The corresponding Poincar\'e map (Fig.~\ref{irc}, 
  right) shows four clearly distinct stripes (black). 
  A resonance such as in the case of the previously studied models A and B is not found.
 
\begin{figure}[htb]
   \centering
    \includegraphics[width=0.63\textwidth]{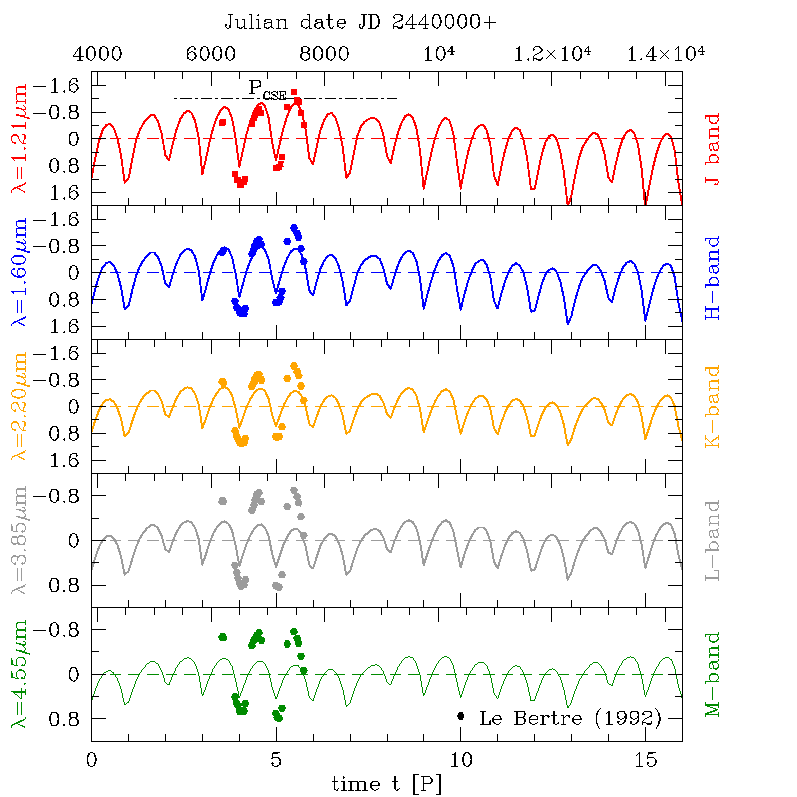}
    \caption{Comparison between calculated (line) \& observed (points) light curves of IRC +10216 at different 
             NIR-wavelengths. The magnitudes are related to the mean magnitude. The CSE-eigenperiod $P_{\mathrm{CSE}}$ 
             is marked with a dashed-dotted line.}
    \label{irci_light}
  \end{figure}

  Figure \ref{irci_light} compares specific observations and synthetic light curves in different NIR bands.
  The stellar pulsation period $P = 650\,\mathrm{d}$ can be seen in all bands. An additional period of six 
  times the stellar pulsation period is also visible (dashed-dotted line). 
  This is the dust-determined eigenperiod of the CSE $P_{\mathrm{CSE}}\approx 3900\,\mathrm{d}$. Due to the 
  short observation period the data cannot reproduce the long envelope period. Nevertheless on can see a 
  continuous upward trend in the observation data (J, H, K-band). This provides a hint of the underlying 
  dust-determined period. 

 \section{Summary}
  We investigated the dynamics of carbon-rich envelopes around AGB stars. We showed that the presence of dust has 
  an important influence on the dynamical behaviour of the entire system. Firstly, it amplifies the momentum 
  coupling between matter and radiation. Secondly, it enforces an additional dynamical behaviour (exterior $\kappa$-
  mechanism) to the dust envelope. The consequence of the coupling is the development of an autonomous envelope 
  dynamics which is in fact excited by the stellar pulsation but caused by the timescales of physical processes 
  in the envelope. This dynamics can be characterised by distinct frequencies which do not need to be identical 
  with the stellar pulsation frequency. Depending on the excitation the CSEs show distinct dynamics regimes.
  We have found a resonance condition which relates the stellar pulsation period to the envelope modes. 
  Besides the pulsation period, light curves exhibit timescales which are determined by the dust.

\end{document}